\documentclass[twocolumn,pra,showpacs,preprintnumbers,amsmath,amssymb]{revtex4-1}

\usepackage{graphicx}
\usepackage{dcolumn}
\usepackage{subfigure}
\usepackage{epsfig}
\usepackage{pstricks,pst-grad}
\usepackage{pslatex} 

\def\ket #1{\vert #1\rangle}
\def\bra #1{\langle #1\vert}
\def\ketbra #1#2{\ket{#1}\!\bra{#2}}
\newcommand{\braket}[2]{\ensuremath{\left\langle#1|#2\right\rangle}}

\newcommand{\w}{\ensuremath{\mathrm{\omega}}}

\renewcommand{\i}{\mathrm{i}}

\newcommand{\Zp}{\ensuremath{\mathbb{Z}_p}}

\newcommand{\Jam}{Jamio\l kowski }

\newtheorem{theorem}{Theorem}

\DeclareMathAlphabet{\mathcal}{OMS}{cmsy}{m}{n}

\begin{document}

\title{Bipartite entangled stabilizer mutually unbiased bases as maximum cliques of Cayley graphs}

\author{Wim van Dam}
\email{vandam@cs.ucsb.edu}
\thanks{}
\affiliation{%
Department of Computer Science, University of California, Santa Barbara, CA 93106, USA\\
Department of Physics, University of California, Santa Barbara, CA 93106, USA
}
\author{Mark Howard}
\email{mhoward@physics.ucsb.edu}
\affiliation{%
Department of Physics, University of California, Santa Barbara, CA 93106, USA
}%
\date{\today}

\begin{abstract}
We examine the existence and structure of particular sets of mutually unbiased bases (MUBs) in bipartite qudit systems. In contrast to well-known power-of-prime MUB constructions, we restrict ourselves to using maximally entangled stabilizer states as MUB vectors. Consequently, these bipartite entangled stabilizer MUBs (BES MUBs) provide no local information, but are sufficient and minimal for decomposing a wide variety of interesting operators including (mixtures of) \Jam states, entanglement witnesses and more. The problem of finding such BES MUBs can be mapped, in a natural way, to that of finding maximum cliques in a family of Cayley graphs. Some relationships with known power-of-prime MUB constructions are discussed, and observables for BES MUBs are given explicitly in terms of Pauli operators.
\end{abstract}

\maketitle

\section{Introduction.}
One of the most important and long-studied tools in quantum information theory is that of mutually unbiased bases (MUBs). Two orthonormal bases $\mathcal{A}=\{\ket{a}\}$ and $\mathcal{B}=\{\ket{b}\}$ in a Hilbert space of dimension $d$ are said to be mutually unbiased when $|\langle a\ket{b}|=1/\sqrt{d}$ i.e. certainty of a measurement outcome in one basis implies complete uncertainty of a measurement outcome in another. This is the finite-dimensional analogue to the complementarity of position and momentum in continuous variable quantum mechanics. Typically, MUBs are most useful in Hilbert spaces, $\mathcal{H}_d$, of prime power dimension ($d=p^k$), for which \emph{complete} sets of MUBs are known to exist and a number of construction methods are available. Ignoring the trace component (which is often known or unimportant), decomposing a $d \times d$ Hermitean operator (e.g. a density matrix) requires $d^2-1$ parameters, which necessitates measuring $d+1$ different observables (since each observable yields $d-1$ independent probabilities). Complete sets of MUBs are sets with $d+1$ orthonormal bases, possessing the desirable properties of being both mutually unbiased with respect to one another and also being minimal in terms of the number of observables required (hence this is considered the optimal tomography set-up \cite{Adamson:2010,Wootters:1989}).

Our work here concerns the construction of MUBs in Hilbert space of dimension $d=p^2$ that are deliberately incomplete in that they contain only $p^2-1$ observables -- insufficient for parameterizing all operators in $\mathcal{H}_{p^2}$, but sufficient and \emph{minimal} for the description of Hermitean operators that are local maximally mixed (LMM) \cite{Baumgartner:2007}. LMM operators, $W$, defined on a bipartite system $\mathcal{H}_{p^2}=\mathbb{C}^p\otimes \mathbb{C}^p$ are those for which $\text{Tr}_1 (W)=\text{Tr}_2 (W)\propto\mathbb{I}$. This class of operators is surprisingly broad. The \Jam isomorphism, for example, tells us that any unital map $\mathcal{E}$ acting on $\mathcal{H}_p$ can be represented by an LMM operator, indicating that this result could potentially be useful for the characterization of noise processes, whilst reducing the number of measurements required (process tomography using a similar construction is discussed in detail in \cite{Scott:2008}). Other scenarios in which the non-local information is of paramount importance include investigation of bipartite entangled and non-local states, and the witnesses \cite{Guhne:2003} and Bell inequalities \cite{Masanes2003} that identify them. As a final example, the motivation for this work came in considering a convenient, minimal basis with which to decompose so-called Clifford witnesses for detecting stabilizer vs. nonstabilizer operations \cite{WvDMH:2010}.

The literature concerning MUBs, constructions and related structures is vast. This field of study seems to originate with Schwinger's construction for unitary operator bases \cite{Schwinger:1960} in 1960, and subsequently Ivonovic's 1981 construction \cite{Ivonovic:1981} for complete MUBs in prime dimensions. Wootters and Fields \cite{Wootters:1989} provided a MUB construction for power-of-prime dimensions $d=p^k$ and showed its optimality for state reconstruction (tomography). A more recent (2002) construction that also works for power-of-prime dimensions is given by Bandyopadhyay \emph{et al.} \cite{Bandyopadhyay2002} and this is framed explicitly in terms of Pauli operators and stabilizer states. Lawrence \emph{et al.} \cite{Lawrence:2002} found a similar construction for multi-qubit systems in the same year. Since then a large number of related results have been published e.g. \cite{Klimov:2007,Sulc:2007,Klimov:2009,Kalev:2009} (also see a recent review article \cite{Durt:2010} and references therein) and a number of interesting connections with combinatorics (e.g. mutually orthogonal Latin squares \cite{Paterek:2009}) and finite geometry \cite{Saniga:2004,Bengtsson:2005,Wootters:2006}. A prominent example of the usefulness of MUBs is their optimality for state or process reconstruction (a recent experimental result \cite{Adamson:2010} shows an improvement over standard techniques by using MUB state tomography). Quantum key distribution schemes \cite{Cerf:2002,Chau:2005} typically rely on MUBs for their security. Another important application of MUBs is their interpretation in terms of finite phase space, leading to a discrete Wigner function; for a particular choice of MUB using stabilizer states, the resulting Wigner function can shed light on the computational power of circuits in the so-called ``Clifford computer" model \cite{Gibbons:2004,Galvao:2005,Cormick:2006,WvDMH:2010}.

Inadvertently, we have rediscovered some results that were previously known in the context of quantum key distribution \cite{Chau:2005}, and in the context of unitary designs \cite{Gross:2007,Scott:2008} (i.e., the Cliffords that we use to create some of our BES-MUBs are known to create a minimal unitary design). Recent work by Planat \cite{Planat:2011} is somewhat related to our current investigation, insofar as it utilizes graph theoretical concepts and stabilizer (Pauli operator) observables to examine the construction of MUBs. Kalev \emph{et al.} \cite{Kalev:2009} investigated MUBs in bipartite systems using sets of commuting Pauli operators, but their work is more focused on complete sets of MUBs for density operators in $\mathcal{H}_{p^2}$.

This work provides an alternative graph-theoretic method (as opposed to unitary designs or finite field constructions) of analyzing MUBs and similar structures in quantum information theory. It is hoped that a combination of the alternative methods outlined here, in addition to those of \cite{Chau:2005,Gross:2007,Scott:2008,Planat:2011} and others, will prove fruitful for further analyses. We show how to create an orthonormal basis of $p^2$ stabilizer states in $\mathcal{H}_{p^2}$, given a matrix $F \in SL(2,\Zp)$. Furthermore, we show that the quantity $\text{Tr}(F_i^{-1} F_j)$ indicates whether the bases corresponding to $F_i$ and $F_j$ are mutually unbiased. This leads naturally to a Cayley graph structure wherein graph vertices are given by the elements of $SL(2,\Zp)$, and edges between vertices correspond to mutual unbiasedness of the corresponding bases. The BES MUBs that we seek are easily shown to be maximum cliques of the Cayley graphs, and for primes up to 11 we can partition $SL(2,\Zp)$ into $p$ distinct (non-overlapping) BES-MUBs. For primes 13 and higher, it is an interesting open question whether such BES-MUBs exist, as a deterministic search for the maximum clique is infeasible. For the related question of minimal unitary designs it has been noted by Chau that subgroups of $SL(2,\Zp)$ of a particular size only exist for primes up to 11, but it is not clear that complete BES-MUBs depend in any way on the existence of such subgroups. The family of Cayley graphs under consideration (defined for all primes $p$) is actually the graph complement of a family of Ramanujan graphs, and we are able to list some general graph-theoretic properties that hold for all values of $p$. In section \ref{sec:defns} we review the necessary background concerning the Clifford group and introduce some graph-theoretical concepts that will be useful in later sections. In section \ref{sec:MUBconstruction} we explicitly give the recipe for constructing BES-MUBs and relate our work to a well-known MUB construction that uses finite field methods. Section \ref{graphprops} further explores the quantities and concepts from graph theory that can be applied to our family of Cayley graphs, and finally, Appendix \ref{AppendixA} provides a description of the MUB observables in terms of stabilizer measurements as well commuting sets of Pauli operators.

\section{Definitions and Useful Results}\label{sec:defns}
\subsection{Relevant finite groups and their properties}
The finite-dimensional analogues of position and momentum operators are denoted by $X$ and $Z$, arbitrary products of which are called displacement operators $D$, indexed by a vector $u=(u_1,u_2)\in \Zp^2$:
\begin{align}
&X\ket{j}=\ket{j+1} \quad Z\ket{j}=\omega^j\ket{j} \quad \left(\omega=e^{2\pi\i/p}\right)\\
&D_u=\tau^{u_1 u_2} X^{u_1} Z^{u_2}\quad \tau=e^{(p+1)\pi\i/p}.
\end{align}
The Weyl-Heisenberg group (or generalized Pauli group) for a single qupit is given by
\begin{align}
\mathcal{G}_p=\left\{\tau^c D_u \vert u \in \Zp^2, c \in \Zp\right\}.
\end{align}
The set of unitary operators that map the Pauli group onto itself under conjugation is called the Clifford group (sometimes called the Jacobi group):
\begin{align*}
 \mathcal{C}_p=\{C\in U(p)\vert U \mathcal{G}_p U^\dag =\mathcal{G}_p\}.
\end{align*}
 The fact that every Clifford operation in dimension $p$ can be associated with a matrix $F \in SL(2,\Zp)$ in addition to a vector $u \in \Zp^2$ results from the isomorphism
\begin{align}
\mathcal{C}_p \cong SL(2,\Zp) \ltimes \Zp^2,
\end{align}
established by Appleby \cite{Appleby:arxiv09}, where $\mathcal{C}$ is the Clifford group. If we specify the elements of $F$ and $u$ as
\begin{align}
F=\left(
    \begin{array}{cc}
      \alpha & \beta \\
      \gamma & \delta \\
    \end{array}
  \right)\in SL(2,\Zp)  \qquad u=\left(
                                 \begin{array}{c}
                                   u_1 \\
                                   u_2 \\
                                 \end{array}
                               \right)\in \Zp^2 \label{Fudef}
\end{align}
then Appleby provides an explicit description of the unitary matrix $\protect{C_{(F\vert u)} \in  \mathcal{C}_p}$ in terms of these elements i.e.,
\begin{align}
&C_{(F\vert u)}=D_u U_F \\
&U_F=\begin{cases}
\frac{1}{\sqrt{p}}\sum_{j,k=0}^{p-1}\tau^{\beta^{-1}\left(\alpha k^2-2 j k +\delta j^2\right)}\ket{j}\bra{k}\quad &\beta\neq0\\
\sum_{k=0}^{p-1} \tau^{\alpha \gamma k^2} \ket{\alpha k}\bra{k}\quad &\beta=0.
\end{cases}
\end{align}


Note how composition and inverses can be represented in this notation \cite{Scott:042203}
\begin{align}
C_{(F\vert u)} C_{(K \vert v)}=C_{(FK\vert u+Fv)}\label{composition}\\
C_{(F\vert u)}^{-1}=C_{(F\vert u)}^\dag = C_{(F^{-1} \vert -F^{-1}u)}\label{inverse}
\end{align}

We will have need to relate the matrix trace $\text{Tr}\left(C_{(F\vert u)}\right)$ to the matrix trace $\text{Tr}\left(F\right)$ modulo $p$:
\begin{align}
&\lvert\text{Tr}\left(C_{(F\vert u)}\right)\rvert = \begin{cases} \in\{0,\sqrt{p},p\} &\text{ if }\text{Tr}(F)= 2 \\
1 &\text{ if }\text{Tr}(F)\neq 2 \end{cases}\label{tracerelations}
\end{align}

To see why this is so we must define the Legendre Symbol
\begin{align*}
\ell_p(x) =
\begin{cases}
\;\;\,1\;\;\, \text{ if } &x \text{ is a quadratic residue} \pmod{p}  \\
-1\;\, \text{ if } &x \text{ is a quadratic non-residue} \pmod{p}\\
\;\;\,0\;\;\, \text{ if } &x \equiv 0 \pmod{p}.
\end{cases}
\end{align*}
and quote a result from Appleby \cite{Appleby:arxiv09}
\begin{align}
(\text{ Case 1: }\qquad\beta =0 \Rightarrow \alpha\neq 0\ )\hspace{1cm} \nonumber \\
\lvert \text{Tr}\left(C_{(F\vert u)}\right) \rvert = \begin{cases} |\ell_p(\alpha)|=1 \quad &(\text{Tr}(F)\neq 2) \\
|\ell_p(\gamma)| \sqrt{p} \delta_{u_1,0}\quad &(\text{Tr}(F)= 2, \gamma \neq 0)\\
p \delta_{u_1,0}\delta_{u_2,0}\quad &(\text{Tr}(F)= 2, \gamma =0)
\end{cases}\label{tracerelations1} \\
(\text{ Case 2: }\qquad\beta \neq 0 \ )\hspace{2cm} \nonumber\\
\lvert \text{Tr}\left(C_{(F\vert u)}\right) \rvert = \begin{cases} |\ell_p(\text{Tr}(F)- 2)|=1 \quad &(\text{Tr}(F)\neq 2)\\
|\ell_p(-\beta)| \sqrt{p} \delta_{u_2,\beta^{-1}(1-\alpha)u_1 }\quad &(\text{Tr}(F)= 2)
\end{cases}\label{tracerelations2}
\end{align}

Finally, we note some important facts regarding the structure of the group $ SL(2,\Zp) $. A minimal set of generators is e.g.
\begin{align}
SL(2,\Zp) =\left\langle \left(
                          \begin{array}{cc}
                            1 & 1 \\
                            0 & 1 \\
                          \end{array}
                        \right),\left(
                          \begin{array}{cc}
                            1 & 0 \\
                            1 & 1 \\
                          \end{array}
                        \right) \right \rangle
\end{align}
It has order $\protect{|SL(2,\Zp)|=p(p^2-1)}$ and can be partitioned into $p+4$ conjugacy classes \cite{appleby:012102}, each of which has constant trace. If we partition $ SL(2,\Zp) $ by the matrix trace of its elements, $\text{Tr}(F)$, we see the following
\begin{align}
\Big|\left\{F|\ell_p\left((\text{Tr}(F))^2-4\right)=1\right\}\Big|&=&p(p+1)\\
\Big|\left\{F|\ell_p\left((\text{Tr}(F))^2-4\right)=-1\right\}\Big|&=&p(p-1)\\
\Big|\left\{F|\ell_p\left((\text{Tr}(F))^2-4\right)=0\right\}\Big|&=&p^2\hspace{0.5cm}
\end{align}
The final sets $\left\{F|\text{Tr}(F)=2\right\}$ and $\left\{F|\text{Tr}(F)=-2\right\}$ are each comprised of three conjugacy classes. Many of these facts will be used in subsequent sections, particulary section \ref{graphprops} concerning graph-theoretical properties of Cayley graphs that are relevant to the construction of BES MUBs.

\begin{figure}
\begin{center}
\subfigure{\epsfig{file=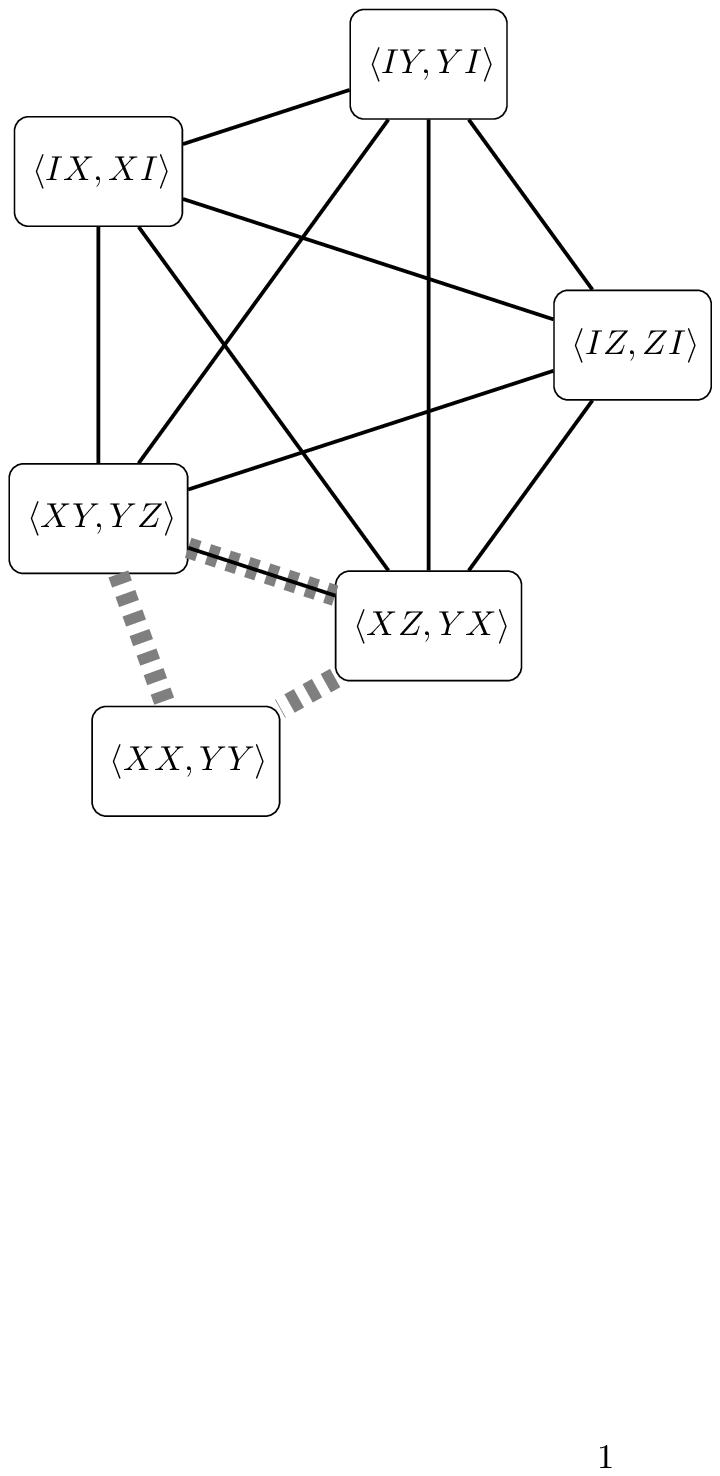,bb=120 250 350 550 ,clip= ,height=70mm}}{{(a)}}%
\vspace{2mm}\\
\subfigure{\epsfig{file=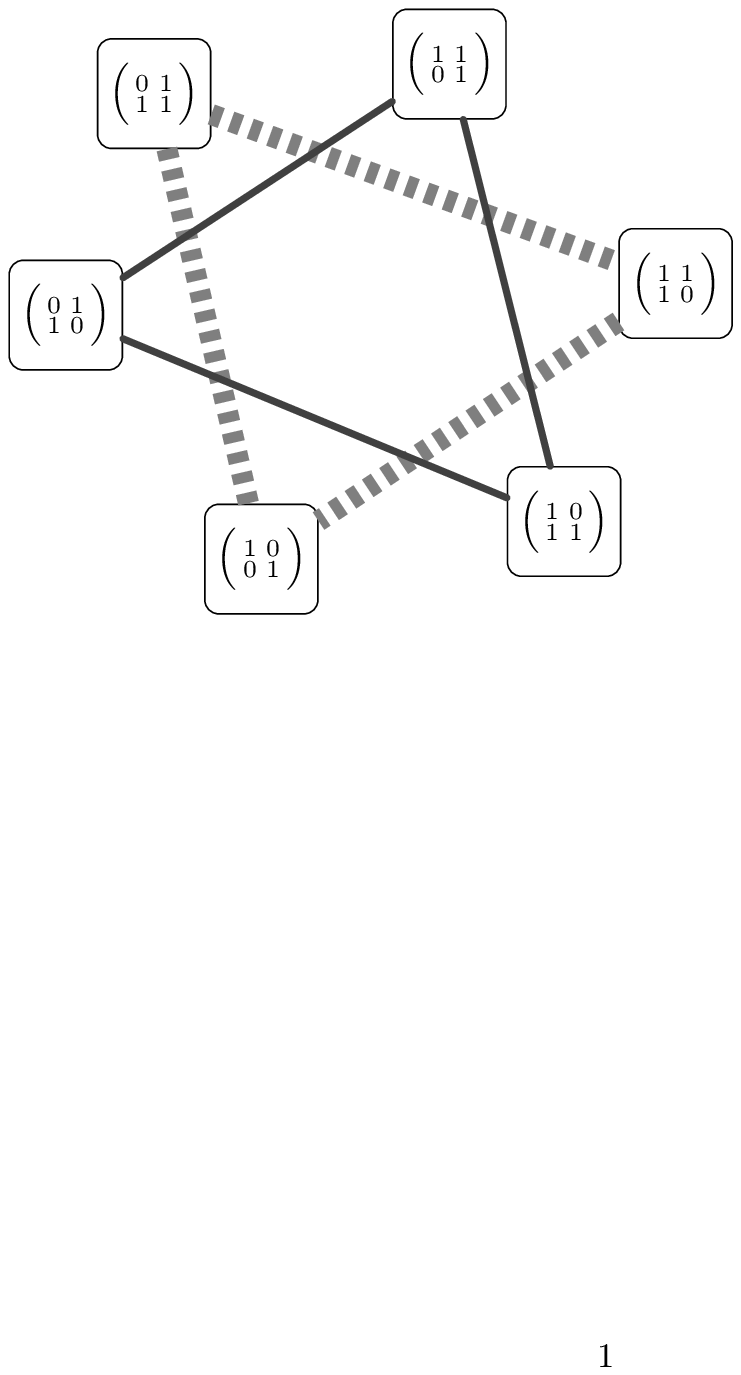,bb=130 250 350 490 ,clip= ,height=50mm}}{{(b)}}
\end{center}
\caption{\label{GraphFig}%
 {MUB structure:} (a) MUBs in dimension $2^2$. Each box represents a two-qubit stabilizer state $\rho=\frac{1}{4}\sum_{s\in \mathcal{S}} s$ where $\mathcal{S}$ is the abelian subgroup generated by the Pauli operators contained in $\langle \cdot \rangle$. Varying the signs of the generators creates a complete orthonormal basis from each representative pair. Lines between boxes indicate that the overlap between two states is $\text{Tr}(\rho_a \rho_b)=\frac{1}{4}$. The solid lines depict the complete graph on 5 vertices, $K_5$, and this corresponds to a complete MUB on this Hilbert space. The dashed lines depict a triangle, $K_3$, which forms a BES MUB. (b) Two different BES MUBs (solid and dashed complete graphs $K_3$) that partition $SL(2,\mathbb{Z}_2)$, where each $2 \times 2$ matrix $F$ corresponds to the \Jam state $(I\otimes C_{(F\vert 0)}) \sum_{j=0}^{j=1} \ket{jj}/\sqrt{2}$. Adjacent vertices $\protect{F_1,F_2 \in SL(2,\Zp)}$ satisfy $\text{Tr}(F^{-1} F_2)\neq 2$, which in terms of the corresponding density matrices implies $\text{Tr}(\rho_{{}_{(F_1)}} \rho_{{}_{(F_2)}})=\frac{1}{4}.$}
\end{figure}

\subsection{Graphs: Cayley Graphs and Maximum Cliques}

We review some relevant notation and properties of graphs that can be found in any standard reference (e.g, \cite{vanLint:1975}). An undirected Cayley graph $\Gamma(G,T)$ with an associated finite group $G$ and set $T \subset G$, is the graph whose vertices are the elements of $G$ and whose set of edges is $\protect{\{ g_1\sim g_2 |g_1^{-1}g_2 \in T\}}$. We must have $I \not\in T$ and $T^{-1}=T$. The resulting graph $\Gamma(G,T)$ is regular i.e. each vertex has degree $|T|$, and the number of (undirected) edges is given by $\frac{1}{2}|G||T|$ .
A complete graph of order $n$, denoted $K_n$, is a graph with $n$ vertices, each of which is adjacent to every other vertex (see Fig \ref{GraphFig} (a) for an example $K_5$). A subgraph, $\Gamma^{\ \prime}$, of $\Gamma$, is a graph whose vertices form a subset of the vertices of $\Gamma$ and the adjacency relation is inherited from $\Gamma$. A clique of $\Gamma$ is a complete subgraph of $\Gamma$, where the size of the clique is given by the number of vertices in this subgraph. The largest possible clique (not necessarily unique) contained in $\Gamma$ is a maximum clique, the size of which is usually denoted $\omega(\Gamma)$. We discuss graph-theoretic properties, and what they say about the problem at hand, in more detail in Section.~\ref{graphprops}.

\section{Construction of the restricted MUB}\label{sec:MUBconstruction}

The goal is to create a set of states, $\mathcal{S}$,  of size $|\mathcal{S}|=p^2(p^2-1)$ that is partitioned into $p^2-1$ subsets, where each subset, containing $p^2$ states, forms an orthonormal basis. Labeling the basis with a supercript and the individual states within a basis using a subscript we have
\begin{align*}
\mathcal{S}=\{\ket{\psi_1^1}\ldots \ket{\psi_j^k}\ldots \ket{\psi_{p^2}^{p^2-1}}\}.
\end{align*}
This is a mutually unbiased basis if
\begin{eqnarray*}
|\langle \psi_j^k \vert \psi_m^n\rangle|=\frac{1}{p}(1-\delta_{k,n})+\delta_{k,n}\delta_{j,m}.
\end{eqnarray*}
The $ \ket{\psi_j^k}$ of the set $\mathcal{S}$, that comprises our bipartite entangled stabilizer MUB (BES MUB), will be maximally entangled stabilizer states -- formed by applying a Clifford operation, $C$, to one half of a maximally entangled state
\begin{eqnarray*}
\ket{J_C}=(I\otimes C) \sum_{j=0}^{p-1} \frac{\ket{jj}}{\sqrt{p}}.
\end{eqnarray*}

The overlap $|\langle J_{C_m}\vert J_{C_n}\rangle|$ between any two such states is given by
\begin{align*}
|\langle J_{C_m}\vert J_{C_n}\rangle|=\frac{1}{p}|\text{Tr}(C_m^\dag C_n)|.
\end{align*}


Using the notation we have previously described, it is easy to show using Eq.s~$\protect{\ref{composition} - \ref{tracerelations}}$ that
\begin{align}
\text{if }\quad &\text{Tr}(F^{-1} K) \neq 2 \nonumber\\
\text{then }\quad \Big\lvert &\text{Tr}\left[\left(C_{(F\vert u)}\right)^\dag \left(C_{(K \vert v)}\right) \right]\Big\rvert=1 \quad \forall u,v \in \Zp^2
\end{align}
i.e., a pair of matrices $F,K \in SL(2,\Zp)$ satisfying $\protect{\text{Tr}(F^{-1} K)\neq 2}$ defines a pair of mutually unbiased basis.  Since the subspace under consideration has dimension $\protect{(p^2-1)^2}$, and since each basis contains $p^2-1$ independent states, we require a total of $\protect{p^2-1}$ matrices $\protect{F_i \in SL(2,\Zp)}$, satisfying, pairwise, $\protect{\text{Tr}(F_i^{-1} F_j)\neq 2}$, in order to create the BES MUB.


Define
\begin{align}
&G=SL(2,\Zp) &|G|=p(p^2-1) \label{graphdef}\\
&T=\{F\in SL(2,\Zp)| \text{Tr}(F)\neq 2\} \quad &|T|=|G|-p^2 \nonumber
\end{align}
then the Cayley graph $\Gamma(G,T)$ has the property that two vertices $F_i$ and $F_j$ are adjacent if and only if $\protect{\text{Tr}(F_i^{-1} F_j)\neq 2}$. A clique of size $p^2-1$, if it exists, immediately gives the desired complete BES MUB by the preceding discussion. Furthermore, a clique of size $\protect{p^2-1}$ must be a maximum clique since the dimension of the Hilbert space for local maximally mixed operators is $\protect{(p^2-1)^2}$.

\begin{theorem}\label{BEScliques}
A pair of matrices $F_1,F_2 \in SL(2,\Zp)$ satisfying $\protect{\text{Tr}(F_1^{-1} F_2)\neq 2}$ defines a pair of mutually unbiased bases in $\mathcal{H}_{p^2}=\mathbb{C}^p\otimes \mathbb{C}^p$ (via the relationship between $SL(2,\Zp)$ and the Clifford group). A set of matrices $\mathcal{F}=\{F_i\}$, of order $\protect{|\mathcal{F}|=p^2-1}$, such that pairwise $\protect{\text{Tr}(F_i^{-1} F_j)\neq 2  \pmod{p} }$, defines \textit{(i)} a complete bipartite entangled stabilizer MUB \textit{(ii)} a maximum clique of the Cayley graph defined in Eq.~(\ref{graphdef}).
\end{theorem}

One can check using a computer algebra system \cite{GAP4,Hibbard:1998} that the following subgroups $\protect{H_p \leq SL(2,\Zp)}$ have order $\protect{|H_p|=p^2-1}$, and every pair of elements $\protect{F_i, F_j \in H_p}$ satisfies $\protect{\text{Tr}(F_i^{-1} F_j)\neq 2}$ (i.e. these subgroups provide complete BES MUBs).
\begin{align}
p=3: \quad &H_3=\left\langle \left(
                          \begin{array}{cc}
                            0 & 1 \\
                            2 & 0 \\
                          \end{array}
                        \right),\left(
                          \begin{array}{cc}
                            1 & 1 \\
                            1 & 2 \\
                          \end{array}
                        \right) \right \rangle \\
                        p=5: \quad &H_5=\left\langle \left(
                          \begin{array}{cc}
                            0 & 2 \\
                            2 & 0 \\
                          \end{array}
                        \right),\left(
                          \begin{array}{cc}
                            1 & 1 \\
                            2 & 3 \\
                          \end{array}
                        \right) \right \rangle \\
                        p=7: \quad &H_7=\left\langle \left(
                          \begin{array}{cc}
                            0 & 2 \\
                            3 & 0 \\
                          \end{array}
                        \right),\left(
                          \begin{array}{cc}
                            1 & 1 \\
                            4 & 5 \\
                          \end{array}
                        \right) \right \rangle \\
                        p=11: \quad &H_{11}=\left\langle \left(
                          \begin{array}{cc}
                            0 & 1 \\
                            10 & 0 \\
                          \end{array}
                        \right),\left(
                          \begin{array}{cc}
                            0 & 4 \\
                            8 & 10 \\
                          \end{array}
                        \right) \right \rangle
\end{align}

In fact for every prime dimension $p\leq 11$ we can partition $SL(2,\Zp)$ by using $p$ distinct max-cliques of size $p^2-1$. For odd primes it suffices to consider the left cosets of $H_p$ in  $SL(2,\Zp)$ where
\begin{align}
F_t=\left(
      \begin{array}{cc}
        1 & 0 \\
        t & 1 \\
      \end{array}
    \right) \quad t \in \Zp
\end{align}
are the left coset representatives. For $p=13$ and higher, we were unable to find cliques saturating the upper bound of $\protect{p^2-1}$. It is known that, for any primes $p\geq 13$, there does not exist a subgroup $H_p$ of size $\protect{|H_p|=p^2-1}$ \cite{Dickson:1958}, but we are unaware of any proof that cliques of size $p^2-1$ (i.e. complete BES MUBs for $p$-dimensional systems) necessarily depend on this subgroup structure. A deterministic search for a clique of size 168 in the $\Gamma(G,T)$ graph for $p=13$ is infeasible, given the computational complexity of the max-clique problem. A heuristic search was able to find a clique of size 158, however.

By adapting a well-known power-of-prime construction for complete MUBs (Bandyopadhyay et al. \cite{Bandyopadhyay2002}) we can show that the size of the largest clique satisfies
\begin{align} \label{cliquelowerbound}
\omega(\Gamma)\geq p(p-1) \quad \forall p
\end{align}
To be specific, Section 4.3.1 of \cite{Bandyopadhyay2002} describes the construction of a complete set of MUBs for dimensions $p^2$. In their notation, this amounts to finding a set of  $p^2$ $2 \times 2$ symmetric matrices $\{A\}$ such that $\det(A_j-A_k)\neq 0$. Suitable sets of matrices are parameterized by two elements $s,t \in \Zp$ via
\begin{align}
\{A\}=\Big\{\left(
        \begin{array}{cc}
          a & b \\
          b & sa+tb \\
        \end{array}
      \right),\quad \forall\ a,b \ \in \Zp \Big\} \label{cliquelb}
\end{align}
A little thought reveals that every $A$ with non-zero off diagonal element $b$ can be related one-to-one with a matrix $F \in SL(2,\Zp)$, where $F$ has a non-zero element $\beta$ ($F$ defined as per Eq.~\eqref{Fudef}),
\begin{align*}
F(a,b,s,t)=\left(
      \begin{array}{cc}
        -ab^{-1} & -b^{-1} \\
        b-a^2b^{-1}s-at & -ab^{-1}s-t \\
      \end{array}
    \right).
\end{align*}
One can check that the $\det(A_j-A_k)\neq 0$ condition translates to $\protect{\text{Tr}(F_j^{-1} F_k)\neq 2}$, as one would expect. In this way we can create a set of $p(p-1)$ matrices $F \in SL(2,\Zp)$ that form a clique in our Cayley graph $\Gamma(G,T)$. In general, sets of matrices formed this way cannot be extended with an additional $p-1$ matrices $F_k$ (having $\beta_k=0$) to form a complete BES MUB  i.e., they form (part of) a \emph{maximal}, but not maximum, clique in $\Gamma(G,T)$. However, we can often slightly improve upon the lower bound e.g., we can construct cliques of size $p(p-1)+2$ for primes up 17. A consequence of Eq.~(\ref{cliquelowerbound}) is that the fraction of pairs $(F_i,F_j)$ that do not define mutually unbiased bases, out of the total number of such pairs $(F_i,F_j)$, vanishes as $p \rightarrow \infty$. In Appendix~\ref{AppendixA}, we explicitly give the observables involved in these BES MUBs in terms of tensor products of Pauli operators.

\section{Some graph-theoretic properties of these Cayley graphs}\label{graphprops}

In this section we further investigate the graph-theoretical properties of the family of Cayley graphs that were previously shown to be closely related to BES MUBs. Without loss of generality, the elements $F \in SL(2,\Zp)$ can be ordered lexicographically by the vectors constituting the rows of the matrix $F$ i.e.
\begin{align*}
\{F_i\}=\Big\{F_1=\left(
      \begin{array}{cc}
        0 & 1 \\
        -1 & 0 \\
      \end{array}
    \right),F_2=\left(
      \begin{array}{cc}
        0 & 1 \\
        -1 & 1 \\
      \end{array}
    \right),\ldots\\
    \ldots F_{p(p^2-1)}=\left(
      \begin{array}{cc}
        -1 & -1 \\
        -1 & -2 \\
      \end{array}
    \right)\Big\}=\Big\{\left(
      \begin{array}{cc}
        \alpha_i & \beta_i \\
        \gamma_i & \delta_i \\
      \end{array}
    \right)\Big\}.
\end{align*}
It is easy to see that (i) there are $p^2-1$ possibilities for $(\alpha,\ \beta)$; (ii) each such $(\alpha,\ \beta)$ in turn allows for $p$ possible $(\gamma,\ \delta)$. Any two elements $F_i,\ F_j$, for which $(\alpha_i,\ \beta_i)=(\alpha_j,\ \beta_j)$, cannot be connected by an edge since
\begin{align}
\text{Tr}\Big(\left(
      \begin{array}{cc}
        \alpha & \beta \\
        \gamma_i & \delta_i \\
      \end{array}
    \right)^{-1}\left(
      \begin{array}{cc}
       \alpha & \beta\\
        \gamma_j & \delta_j \\
      \end{array}
    \right)\Big)=\det F_i+\det F_j=2.
\label{colorings}
\end{align}
The so-called vertex coloring problem for graphs involves assigning a label (color) to every vertex of the graph, such that adjacent vertices cannot be assigned the same color. The minimum number of colors required to do this is the chromatic number, denoted $\chi(\Gamma)$. It is a basic fact \cite{vanLint:1975} that the chromatic number of a graph is bounded below by the clique number i.e. $\omega(\Gamma)\leq \chi(\Gamma)$.
The discussion leading to Eq.~\ref{colorings} immediately implies that a $p^2-1$ coloring of the Cayley graph $\Gamma(G,T)$ is possible: assign the same color to two vertices $F_i,\ F_j$ if and only if $(\alpha_i,\ \beta_i)=(\alpha_j,\ \beta_j)$. Since the chromatic number $\chi$ is bounded below by the clique number $\omega(\Gamma)$, we know that this coloring is minimal for primes 2 to 11. Hence
\begin{align*}
&\omega(\Gamma) = \chi(\Gamma)= p^2-1& \quad &p\in \{ 2,3,5,7,11\} \hfill \\
&\omega(\Gamma) \leq \chi(\Gamma)\leq p^2-1& \quad \forall &p  \hfill
\end{align*}
Note that the upper bound $\omega(\Gamma) \leq p^2-1$ is a graph-theoretical inequality that confirms the geometrical argument preceding Theorem \ref{BEScliques} i.e., the number of BES mutually unbiased bases that can fit in a Hilbert space $\mathcal{H}_{p^2}=\mathbb{C}^p\otimes \mathbb{C}^p$ is at most $p^2-1$.

A concept closely related to cliques and colorings is that of independence. An independent set of a graph is a set of vertices, no two of which are adjacent. A maximum independent set is the largest such set (not necessarily unique) that can be found in the graph, and the independence number, $\alpha(\Gamma)$, of a graph is the size of this maximum independent set. The discussion preceding Eq.~(\ref{colorings}) can equally well be interpreted as providing a lower bound on the independence number of $\Gamma(G,T)$; there are $p^2-1$ independent sets of size $p$, wherein two elements $F_i,\ F_j$ that satisfy $(\alpha_i,\ \beta_i)=(\alpha_j,\ \beta_j)$ are pairwise non-adjacent, hence
\begin{align}
\alpha(\Gamma)\geq p \qquad \forall p \label{alphalb}
\end{align}
The physical interpretation of this is that we can always find a set of $p$ bases such that, pairwise, no two are mutually unbiased with respect to each other.

The adjacency matrix of a graph $\Gamma$ with $n$ vertices is an $n \times n$ matrix $A[\Gamma]$ with elements $A_{i,j}=1$ if vertices $i$ and $j$ are adjacent, and $A_{i,j}=0$ otherwise. Knowledge of the spectrum of an adjacency matrix often allows us to find, or bound, many quantities of interest. We denote the spectrum of the $p(p^2-1) \times p(p^2-1)$ adjacency matrices $A[\Gamma(G,T)]$ as $\{\lambda^{m_0}_0,\lambda^{m_1}_1,\lambda^{m_2}_2,\lambda^{m_3}_3\}$ where $m_i$ denotes the multiplicity of $\lambda_i$. The complement of a graph $\Gamma$, denoted $\overline{\Gamma}$, is the graph with same vertex set as $\Gamma$, but where two vertices are adjacent in $\overline{\Gamma}$ if and only if they are not adjacent in  $\Gamma$. 
The spectrum of a graph and its complement can be related in a simple way for the case of regular graphs (the case we deal with in this work), as the following theorem demonstrates.

\begin{theorem}\label{Brouwer}
\emph{(Brouwer and Haemers \cite{Brouwer:})} Suppose $\Gamma$ is a $k$-regular graph on $n$ vertices with 4 distinct (adjacency) eigenvalues $\{k=\lambda_0>\lambda_1>\lambda_2>\lambda_3\}$. If, in addition, both $\Gamma$ and its complement, $\overline{\Gamma}$, are connected, then $\overline{\Gamma}$ also has 4 distinct eigenvalues, $\{n-k-1>-\lambda_3-1>-\lambda_2-1>-\lambda_1-1\}$.
\end{theorem}

%


The Cayley graphs we studied, defined in Eq.~\eqref{graphdef}, are actually the graph complement of a well known family of graphs (that form a family of Ramanujan graphs, amongst other interesting properties), whose spectrum is known exactly.
\begin{theorem}\label{Lubotzky}
\emph{(Lubotzky \cite{Lubotzky:1994})} Let $G=SL(2,\Zp)$, and let $T$ (i.e., the connection set for the Cayley graph) be the union of the conjugacy classes $c_1$ and $c_{\nu}$ of the elements
\begin{align*}
\left(
      \begin{array}{cc}
        1 & 0 \\
        1 & 1 \\
      \end{array}
    \right), \quad \left(
      \begin{array}{cc}
        1 & 0 \\
        \nu & 1 \\
      \end{array}
    \right),
\end{align*}
where $\nu$ is a generator of the cyclic group $\Zp^*=\Zp/\{0\}$.
Then $T=\protect{\{F\in SL(2,\Zp)|\text{Tr}(F)=2,F\neq I\}}$, $|T|=p^2-1$ and the spectrum of the corresponding Cayley graph $A[\Gamma(G,T)]$ denoted $\{\lambda^{m_0}_0,\lambda^{m_1}_1,\lambda^{m_2}_2,\lambda^{m_3}_3\}$ , is
\begin{align*}
&\lambda_0 &=&&\ &  p^2-1,&\quad& &\ & m_0 &=&&\ & 1  \\
&\lambda_1 &=&&\ &  p-1 , &\quad& &\ & m_1 &=&&\ & (p-2)(p+1)^2/2\\
&\lambda_2 &=&&\ & 0    , &\quad& &\ & m_2 &=&&\ & p^2\\
&\lambda_3 &=&&\ & -(p+1),&\quad& &\ & m_3 &=&&\ &  p(p-1)^2/2\\
\end{align*}
\end{theorem}

Combining the two preceding theorems (connectedness is obviously satisfied by our Cayley graphs) allows us to completely characterize the spectrum of the canonical Cayley graph Eq.~\eqref{graphdef} that we used to search for BES MUBs.
\begin{theorem}\label{MUBgraphspectrum}
(Spectrum of graphs defined in Eq.~(\ref{graphdef})) Let $G=SL(2,\Zp)$ and $T=\protect{\{F\in SL(2,\Zp)| \text{Tr}(F)\neq 2\}}$. Then $|T|=|G|-p^2$ and the spectrum of $A[\Gamma(G,T)]$ denoted $\{\lambda^{m_0}_0,\lambda^{m_1}_1,\lambda^{m_2}_2,\lambda^{m_3}_3\}$ is
\begin{align*}
&\lambda_0&=&&\ &p(p^2-1)-p^2 &\ & &\ & m_0 &=&&\ & 1  \\
&\lambda_1&=&&\ &p            &\ & &\ & m_1 &=&&\ &  p(p-1)^2/2 \\
&\lambda_2&=&&\ &-1           &\ & &\ & m_2 &=&&\ & p^2\\
&\lambda_3&=&&\ &-p           &\ & &\ & m_3 &=&&\ &  (p-2)(p+1)^2/2
\end{align*}
\end{theorem}

At this point we note that the problem of finding BES MUBs, framed as finding maximum cliques of size $p^2-1$ in the Cayley graph $\Gamma$ defined by Eq.~\eqref{graphdef}, is completely equivalent to finding maximum independent sets of size $p^2-1$ in the complement, $\overline{\Gamma}$, of that graph i.e.,
\begin{align*}
\exists \text{ complete BES MUB } \iff \omega(\Gamma)=p^2-1=\alpha(\overline{\Gamma}).
\end{align*}

%

Unfortunately, it seems that existing spectral lower bounds on the clique number are of little help for the task of proving existence of BES MUBs. Nonetheless, using some well-known spectral bounds we list some implications for the graphs $\Gamma(G,T)$ under consideration. A lower bound on the chromatic number is given by
\begin{align*}
\chi(\Gamma)\geq 1-\frac{\lambda_0}{\lambda_3} = p(p-1),
\end{align*}
which, in conjunction with Eq.~(\ref{colorings}), shows that $\protect{p(p-1) \leq \chi(\Gamma) \leq p^2-1}$. In fact, this lower bound was already implied by Eq.~(\ref{cliquelowerbound}).

For a regular graph, $\Gamma$, on $n$ vertices, Hoffman (unpublished) and  Lov{\'a}sz \cite{Lovasz:1979} proved the formula
\begin{align*}
\alpha(\Gamma)\leq \frac{-n \lambda_{\min}}{\lambda_{\max}-\lambda_{\min}}= \frac{-n \lambda_3}{\lambda_0-\lambda_3}=p+1,
\end{align*}
which, in conjunction with Eq.~(\ref{alphalb}) gives us $\protect{p \leq \alpha(\Gamma) \leq p+1}$. As a final remark on spectral implications, we note that the spectrum exhibited in Thm.~\ref{MUBgraphspectrum} classifies $\Gamma(G,T)$ as a so-called walk-regular graph \cite{Godsil:1980}.

\section{Conclusion}
We have shown how the set of bipartite entangled stabilizer (BES) states can be partitioned into sets of mutually unbiased bases (MUBs), whose span is sufficient and minimal to describe an interesting class of operators that includes (mixtures of) \Jam states, Clifford witnesses \cite{WvDMH:2010} and more. Mutual unbiasedness of two stabilizer orthonormal bases is easily shown to be equivalent to a simple relation on pairs of matrices from $SL(2,\Zp)$. Pairs of matrices satisfying this relation are adjacent vertices on a naturally defined Cayley graph, and the problem of finding complete (optimal) BES MUBs is transformed into that of finding maximum cliques in the Cayley graph. In a different mathematical context, the graph complement of our Cayley graphs are well-studied, and so we can quote, for example, the exact spectrum of the adjacency matrix for all prime values $p$. The most interesting open question is whether such BES-MUBs exist for all primes, or indeed for any primes greater than 11. For the closely related task of finding minimal unitary designs, a discussion by Chau \cite{Chau:2005} (invoking Dickson's theorem on the existence of certain subgroups of $SL(2,\Zp)$) suggests that minimal unitary designs only exist for primes up to 11. It remains to be seen whether the latitude afforded by seeking BES-MUBs, as opposed to subgroups of $SL(2,\Zp)$, allows for construction of optimal BES-MUBs when $p\geq 13$.

%
%

\appendix
\section{Measurement Operators for BES MUBs}\label{AppendixA}
Given a matrix $F \in SL(2,\Zp)$, this defines an orthonormal basis $\mathcal{F}$ in the bipartite Hilbert space $\mathcal{H}_{p^2}$ via
\begin{align}
&\mathcal{F}=\{\ket{J_u^F}, \forall u \in \Zp^2 \},\quad |\braket{J_u^F}{J_v^F}|=\delta_{u,v}\label{Fbasis}\\
&\text{where }\ \ket{J_u^F}=\left(I\otimes C_{(F\vert u)}\right) \sum_{j=0}^{p-1} \frac{\ket{jj}}{\sqrt{p}} \nonumber
\end{align}
We will show how the basis $\mathcal{F}$ can be rewritten in terms of stabilizer measurements, and subsequently how $\mathcal{F}$ can be identified as the simultaneous eigenbasis of a set of $p^2-1$ commuting Pauli operators.

Using so-called symplectic notation, the general form for multi-particle stabilizer operators
with vectors $x=(x_1,x_2,\dots)$ and $z = (z_1,z_2,\dots)$ with $x_i$, $z_i \in \Zp$ is
\begin{align}
P_{(x|z)}=\left(X^{x_1}\otimes X^{x_2}\dots\right) \left(Z^{z_1}\otimes Z^{z_2}\dots\right).
\end{align}
Measuring a two-qupit Pauli operator corresponds to projecting with a rank-$p$ projector, $\Pi$,
\begin{align}
\Pi:=\Pi_{(x_1,x_2|z_1,z_2)[k]}=\frac{1}{p}\big(&I+\w^{-k}P_{(x_1,x_2|z_1,z_2)}+\ldots \nonumber \\
&+\w^{-(p-1)k}(P_{(x_1,x_2|z_1,z_2)})^{p-1}\big)
\end{align}
The product of two appropriately chosen such projectors, $\Pi, \Pi^\prime$, defines a rank-1 operator - a stabilizer state:
\begin{align*}
&\ketbra{\psi}{\psi}=\frac{1}{p^2}\sum_{s\in \mathcal{G}_s} s=\Pi\ \Pi^\prime,
\end{align*}
where $\mathcal{G}_s=\langle g,g^\prime \rangle$ is a subgroup, generated by two commuting Pauli operators $g$ and $g^\prime $, of the group $\mathcal{G}_2=\left\{\w^c P_{(x|z)} \vert x,z \in \Zp^2, c \in \Zp\right\}$. In symplectic notation $g = \omega^{-k}P_{(x_1,x_2|z_1,z_2)}$ and
$g^\prime  = \omega^{-k^\prime} P_{(x^\prime_1,x^\prime_2|z^\prime_1,z^\prime_2)}$ and commutativity of $g$ and $g^\prime$ reduces to
\begin{align*}
\sum_{i=1,2} x_iz_i-x^\prime_iz^\prime_i \equiv 0 \!\mod{p}.
\end{align*}

Given $u=(u_1,u_2) \in \Zp^2$ and $\beta\neq 0$, the following two sets of projectors are equal, up to re-ordering
\begin{align*}
&\forall u:\quad \left\{\ket{J_u^F}\bra{J_u^F}\right\}=\left\{\Pi_{(1,0|\alpha \beta^{-1},-\beta^{-1})[u_1]}\Pi_{(0,1|-\beta^{-1},\beta^{-1}\delta)[u_2]}\right\}.
\end{align*}
When $\beta= 0$, the following two sets of projectors are equal, up to re-ordering
\begin{align*}
&\forall u:\quad \left\{\ket{J_u^F}\bra{J_u^F}\right\}=\left\{\Pi_{(1,\alpha| 0 ,\gamma)[u_1]}\Pi_{(0,0|1,-\delta)[u_2]}\right\}.
\end{align*}

 Many existing constructions for complete MUBs in $\mathcal{H}_{d}$ (with power-of-prime dimension $d$) are based around the partitioning of $d^2-1$ non-identity Pauli operators into $d+1$ classes, each of which contains $d-1$ mutually commuting operators. Each basis within the MUB is then given by the simultaneous eigenbasis of the $d-1$ mutually commuting operators (i.e., each class is associated with exactly one orthonormal basis, for a given partitioning). We can frame the construction of BES MUBs in this language too, with the modification that we are partitioning the set of all weight-two Pauli operators i.e. the subset $\big\{P_{(x_1,x_2|z_1,z_2)}/\{P_{(x_1,0|z_1,0)},P_{(0,x_2|0,z_2)}\}\big\}$ of size $(p^2-1)^2$. With individual classes containing $p^2-1$ operators, there can only be at most $p^2-1$ such classes. It should be clear that a set of $n$ matrices $F\in SL(2,\Zp)$ (satisfying $\protect{\text{Tr}(F_i^{-1} F_j)\neq 2}$) is equivalent to $n$ non-overlapping classes of weight-two Pauli operators, each class containing $p^2-1$ non-identity elements. Recalling Eq.~(\ref{Fbasis}) for the definition of the basis associated with $F$, then the associated class of unitary operators is the subgroup $\mathcal{G}_s=\langle g,g^\prime \rangle$ of $\mathcal{G}_2$. The simultaneous eigenbasis of all $p^2$ Pauli operators in $\mathcal{G}_s$ forms an orthonormal basis.
When $\beta \neq 0$ the class of Pauli operators corresponding to $F$ is given by
\begin{align*}
\mathcal{G}_s(F)=\langle g,g^\prime \rangle:=\langle P_{(1,0|\alpha \beta^{-1},-\beta^{-1})},P_{(0,1|-\beta^{-1},\beta^{-1}\delta)} \rangle.
\end{align*}
When $\beta = 0$ the class of Pauli operators corresponding to $F$ is given by
\begin{align*}
\mathcal{G}_s(F)=\langle g,g^\prime \rangle:=\langle P_{(1,\alpha  | 0,\gamma )},P_{(0,0 | 1,-\delta)} \rangle.
\end{align*}

\end{document}